\magnification=\magstep1
\parskip = 2pt

\newcount\ftnumber
\def\ft#1{\global\advance\ftnumber by 1
          {\baselineskip=13pt \footnote{$^{\the\ftnumber}$}{#1 }}}
\newcount\fnnumber
\def\fn{\global\advance\fnnumber by 1
         $^{(\the\fnnumber)}$}

\def\fr#1/#2{{\textstyle{#1\over#2}}} 

\font\title = cmbx10 scaled 1440 
\font\ss = cmssbx10

\def\lp{\bigl(}
\def\rp{\bigr)}

\def\>{\rangle}
\def\<{\langle}
\def\k#1{|\,#1\>}

\def\x{\otimes}

\def\Sc{\hbox{\ss S}}

\def\oc{\hbox{\ss 1}}
\def\1c{\hbox{\ss 1}} 
 
\def\0c{\hbox{\ss 0}}

\def\Xc{\hbox{\ss X}}
\def\Yc{\hbox{\ss Y}}
\def\Zc{\hbox{\ss Z}}

\def\Uc{\hbox{\ss U}}

\def\Hc{\hbox{\ss H}}

\def\c{\fr1/{\sqrt2}}

\def\x{\otimes}
\def\h{\fr1/2}

\def\ra{\rightarrow}

\def\+{\oplus}

\def\={\equiv}

\newcount\eqnumber

\def\eq(#1){
    \ifx\DRAFT\undefined\def\DRAFT{0}\fi	
    \global\advance\eqnumber by 1%
    \expandafter\xdef\csname !#1\endcsname{\the\eqnumber}%
    \ifnum\number\DRAFT>0%
	\setbox0=\hbox{#1}%
	\wd0=0pt%
	\eqno({\offinterlineskip
	  \vtop{\hbox{\the\eqnumber}\vskip1.5pt\box0}})%
    \else%
	\eqno(\the\eqnumber)%
    \fi%
}
\def\(#1){(\csname !#1\endcsname)}


\def\DRAFT{0}
\baselineskip = 15pt

\font\title = cmb10 scaled 1440

\centerline{{\title Copenhagen Computation:}}
\vskip 5pt
\centerline{{\title How I Learned to Stop Worrying and Love Bohr}}
\vskip 10pt
\centerline{N. David Mermin}

\centerline{Laboratory of Atomic and Solid State Physics}

\centerline{Cornell University, Ithaca, NY 14853-2501}

\vskip 10pt

{\narrower \narrower \baselineskip = 12pt

To celebrate the 60th birthday of Charles H.~Bennett I (1) publicly
announce my referee reports for the original dense coding and
teleportation papers, (2) present a very economical solution to the
Bernstein-Vazirani problem that does not even hint at interference
between multiple universes, and (3) describe how I inadvertently
reinvented the Copenhagen interpretation in the course of constructing
a simple, straightforward, and transparent introduction to quantum
mechanics for computer scientists.

}

\vskip 10pt
\leftline{{\bf 1. Preface: present at the birth}}
\vskip 5pt

David DiVincenzo, Patrick Hayden, and Barbara Terhal [1] have
designated me the ``midwife of teleportation'' in recognition of my
having written a favorable referee's report on the discovery paper [2]
and having advised the editors that the proposed terminology made
sense.  Though this honorific raises vexing biological questions ---
can something with six fathers and no mother be brought forth by a
midwife? --- I accept the title with pride.  As midwife it seemed
appropriate for me to read my referee's report at the Bennett 60th
Birthday Symposium, attended, as it was, by all six fathers.  I
reproduce it here too, since it shows me to have had a taste for
Copenhagen Computation (about which more below) even before Chris
Fuchs [3] got to work on me.

\vskip 10pt
{\narrower \narrower \parindent=0pt \baselineskip=8pt \obeylines
{\tt
Referee's Report: Bennett et al., "Teleporting$\ldots$" LZ4539

\vskip 10pt

This is a charming, readable, thought-provoking paper.  
It presents a novel application of EPR correlations. The 
character of the quantum state (how much is inherent in 
the physical system, how much is a representation of our 
knowledge) is still an extremely elusive notion. This 
novel method for duplicating a quantum state somewhere
else by a combination of quantum correlations and classical
information will become an important one of the 
intellectual tools available to anybody trying to clear 
up this murkiness.

}

}
\vskip 5pt
While hunting down the above report I discovered, to my
amazement that the year before I had also refereed the discovery
paper on dense coding [4].  (I was under the
impression that I had paid no attention whatever to dense coding until
shortly before its deconstruction in 2002 [5].)

\vskip 10pt
{\narrower \narrower \parindent=0pt \baselineskip=8pt \obeylines
{\tt

Bennett and Wiesner, "Communication via one-and 
two-particle$\ldots$" LT4749 
\vskip 10pt
Your question was: Does this qualify as "strikingly 
different" enough to publish?  I have never read anything 
like it, and I have read a lot on EPR, though far from 
everything ever written.  So as far as I know it is 
different.
\vskip 10pt

But strikingly?  The argument is very simple, so 
shouldn't the point be obvious?  After reading the paper 
I put it aside and spent the next week working hard on 
something totally unrelated.  Every now and then I would 
introspect to see if some way of looking at the argument 
had germinated that reduced it to a triviality.  None had.  
Last night I woke up at 3am, fascinated and obsessed with 
it.  Couldn't get back to sleep.  That's my definition of 
"striking".

\vskip 10pt
So I say it's strikingly different and I say publish it.

}

} \vskip 5pt Rereading these old reports reminded me that the myth
that referees relish their power to reject papers is off the mark.
Writing a favorable report for a good paper is sheer pleasure.
Negative reports are no fun at all.

\vskip 10pt
\leftline{{\bf 2. Prologue: Bernstein-Vazirani without parallel universes}}

The Bernstein-Vazirani problem presents one with a black-boxed
subroutine, shown in Figure 1, whose action on $n+1$ qubits is that of
a unitary transformation $\Uc_a$ which takes the computational basis
state $\k x_n\k y_1$ of an $n$-Qbit [6] input register and 1-Qbit
output register into the state $\k x_n\k{y\+ x\cdot a}_1$.  Here $\+$
denotes addition modulo 2, $x\cdot a$ denotes the bitwise modulo 2
inner product of the two $n$-bit numbers $x$ and $a$ ($x\cdot a =
x_{n-1}a_{n-1} \+\cdots\+ x_1a_1\+x_0a_0$), and $a$ is some fixed but
unknown $n$-bit integer with binary expansion $a = a_{n-1}\ldots
a_1a_0$.  The problem is to find the smallest number of invocations of
the black box needed to learn $a$.

\input epsf
\midinsert 
\hrule
\vskip 10 pt
\epsfxsize=2truein
\centerline{\epsfbox{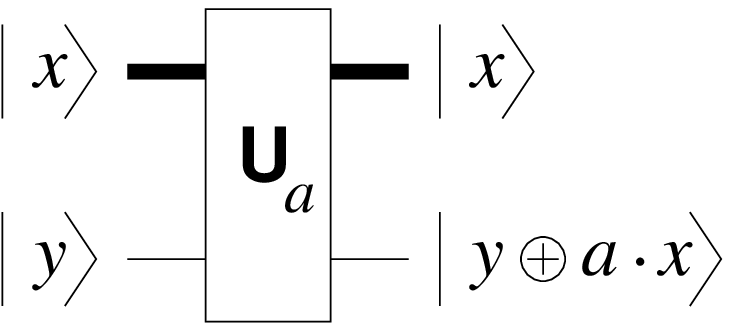}}
\vskip 5pt
{\narrower 

\noindent Figure 1.  The black-boxed Bernstein-Vazirani subroutine
$\Uc_a$. The
heavy wire is the $n$-Qbit input register; the light wire is the
1-Qbit output register. 

}
\vskip 10pt
\hrule
\endinsert 

If the subroutine is applied to $x=2^j$, the output register will be
flipped if and only if $a_j=1$, so a classical computer can determine
$a$ with $n$ calls of the subroutine.  Evidently there is no classical
way to learn $a$ with fewer than $n$ calls, since one needs $n$
independent linear relations among the bits of $a$.  But with a
quantum computer one can find $a$ with just a single call of the
subroutine, whatever the size of $n$.

This remarkable trick is done by applying a Hadamard transformation,
$$\Hc\k0 = \c\lp\k0+\k1\rp,\ \ \ \Hc\k1 = \c\lp\k0-\k1\rp,\eq(had) $$
to every one of the $n+1$ Qbits both before and after the application
of $\Uc_a$, as shown in Figure 2.  If one initializes the input
register to the state $\k0_n$ and the output register to the state
$\k1$, then at the end of this process the input register is
guaranteed to be in the $n$-Qubit state $\k a_n$.  So $a$ can be
learned by measuring each Qbit of the output register in the
computational basis.

\midinsert 
\hrule
\vskip 10 pt
\epsfxsize=2.5 truein
\centerline{\epsfbox{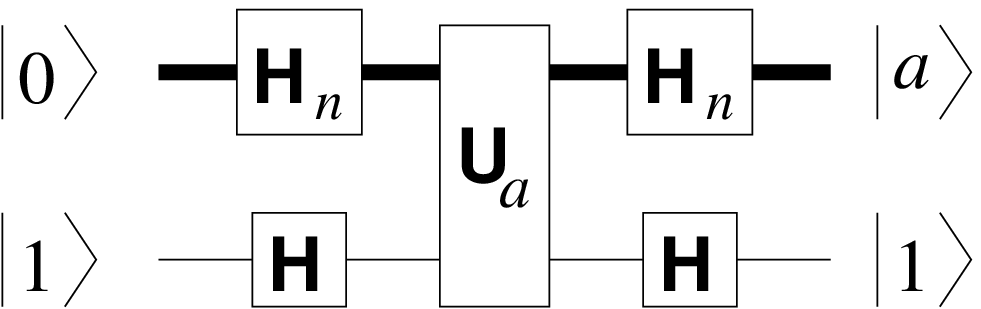}}
\vskip 5pt
{\narrower

\noindent Figure 2. Quantum solution of the Bernstein-Vazirani
problem.  
$\Hc_n$
is an $n$-fold tensor product of 1-Qbit Hadamards.
 
}
\vskip 10pt
\hrule
\endinsert 

The conventional explanation for why this works goes like this: 

(i) Applying Had\-a\-mards to every Qbit of an input register
initially in the $n$-Qbit state $\k0_n$ results in a uniform
superposition of all possible inputs: $$ \Hc_n\k0_n =
2^{-n/2}\sum_{0\leq x<2^n}\k x_n. \eq(inputs)$$ 

(ii) Preparing the output register in the state $\Hc\k1$ converts a
bit-flip into a change of phase (specifically a multiplication by
$-1$). 

(iii) Another application of Hadamards to the input register
after the application of $\Uc_a$ introduces additional $x$-dependent
phases according to the rule $$\Hc_n\k x_n = 2^{-n/2}\sum_{0\leq
z<2^n}(-1)^{x\cdot z}\k z_n. \eq(output)$$ 

(iv) A little arithmetic now reveals that the combined phases lead to
complete destructive interference for every term characterizing the
input register in the final superposition except for the single state
$\k a_n$.

This process is usually described as an application of massive quantum
parallelism followed by destructive interference among all the
unfavorable outcomes.  People with overactive imaginations like to say
that step (i) initializes a computer in each of $2^n$ parallel
universes to each of the $2^n$ possible inputs.  The remaining steps
are cunningly designed to produce destructive interference among all
those $2^n$ universes, in just such a way as to lead in every single
universe to the presence of $a$ in the input register at the end of
the process.
   
There is, however, a much simpler way to understand why the circuit in
Figure 2 behaves as advertised, which offers no hint of this
metaphysical extravaganza.  This approach merely notes that the effect
of Hadamards on the basic 2-Qbit controlled-NOT (cNOT) gate, defined in
Figure 3, is just to interchange the control and target Qbits, as
shown in Figure 4.  This follows from the fact that $\Hc^2 = \oc$ and
$\Hc\Xc\Hc = \Zc$, where $$\Xc\k0 = \k1,\ \ \ \Xc\k1=\k0,\ \ \
\Zc\k0=\k0,\ \ \ \Zc\k1 = -\k1, \eq(Zc)$$ and the fact that
controlled-$\Zc$ is symmetric under interchange of target and control
Qbits.

\vskip 6pt 
\hrule 
\vskip -4pt
\midinsert 
\epsfxsize=2.5truein \centerline{\hskip
25pt \epsfbox{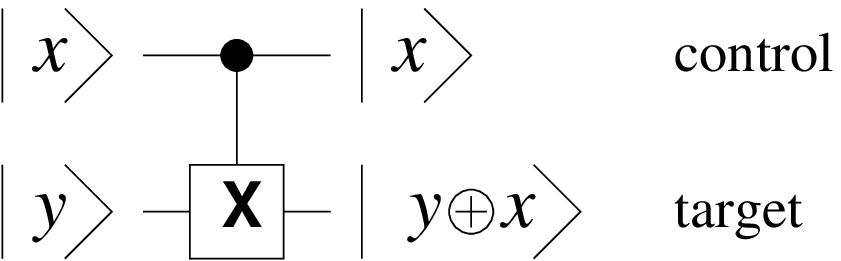}} \vskip 5pt {\narrower

\noindent Figure 3. The 2-Qbit cNOT gate. 

}
\vskip 10pt
\hrule
\endinsert 
\vskip -10pt

\vskip -8pt
\midinsert 
\epsfxsize=2truein
\centerline{\epsfbox{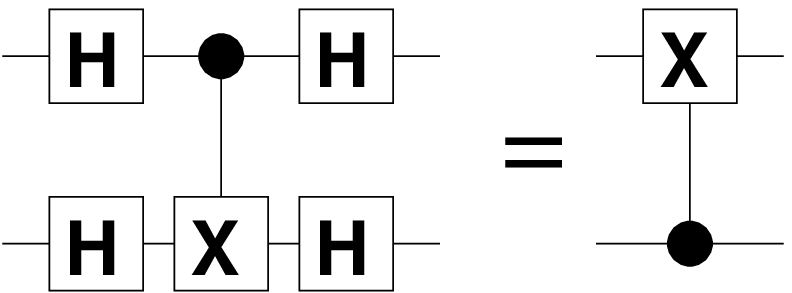}}
\endinsert 
{\narrower
\vskip -6pt
\noindent Figure 4. Hadamards interchange target and control Qbits. 
\vskip 10pt

} \hrule \noindent

The action of $\Uc_a$ shown in Figure 1 is identical to the action
of a collection of cNOT gates --- one for each non-zero bit
of $a$.  They all target the output register, and are controlled by
just those Qbits representing bits of $x$ that correspond to non-zero
bits of $a$.  This is illustrated in Figure 5 for $n=5$ and $a =
11010$.  Since sandwiching cNOT gates between Hadamards
interchanges the control and target Qbits and since $\Hc$ is its own
inverse, the magic of Bernstein-Vazirani follows at once, as shown in
Figure 6.

\midinsert 
\hrule
\vskip 10pt
\epsfxsize=6truein
\centerline{\epsfbox{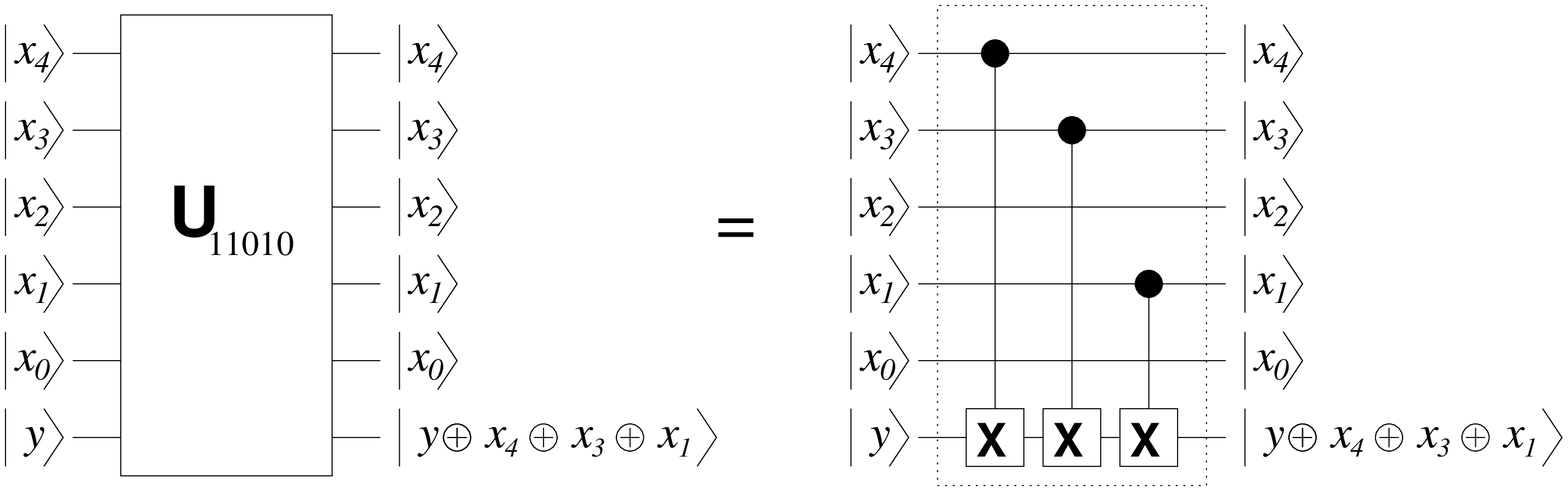}}


{\narrower

\noindent Figure 5.  The black-boxed Bernstein-Vazirani oracle (shown
for the case $n=5$, $a = 11010$) behaves {\it as if\/} it contained a
collection of cNOT gates.  \vskip 10pt } \hrule 
\endinsert

\midinsert \hrule
\vskip 15pt \epsfxsize=3truein \centerline{\epsfbox{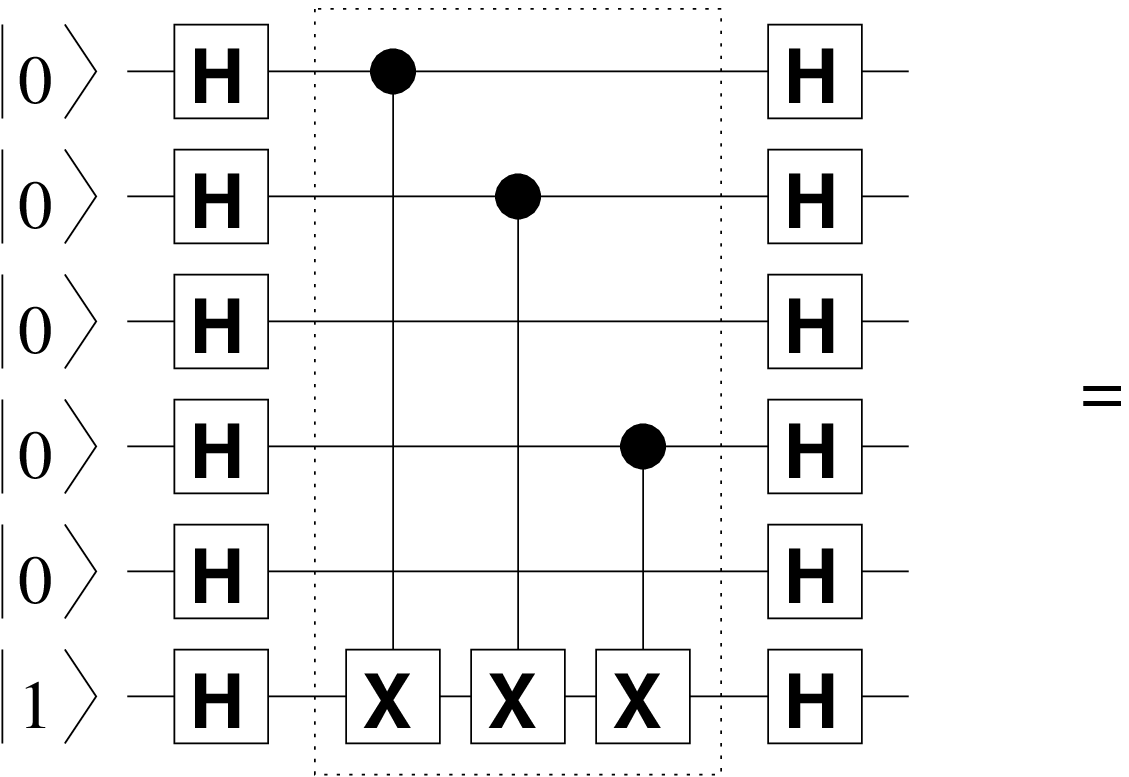}} \vskip
15pt

\epsfxsize=6truein
\centerline{\epsfbox{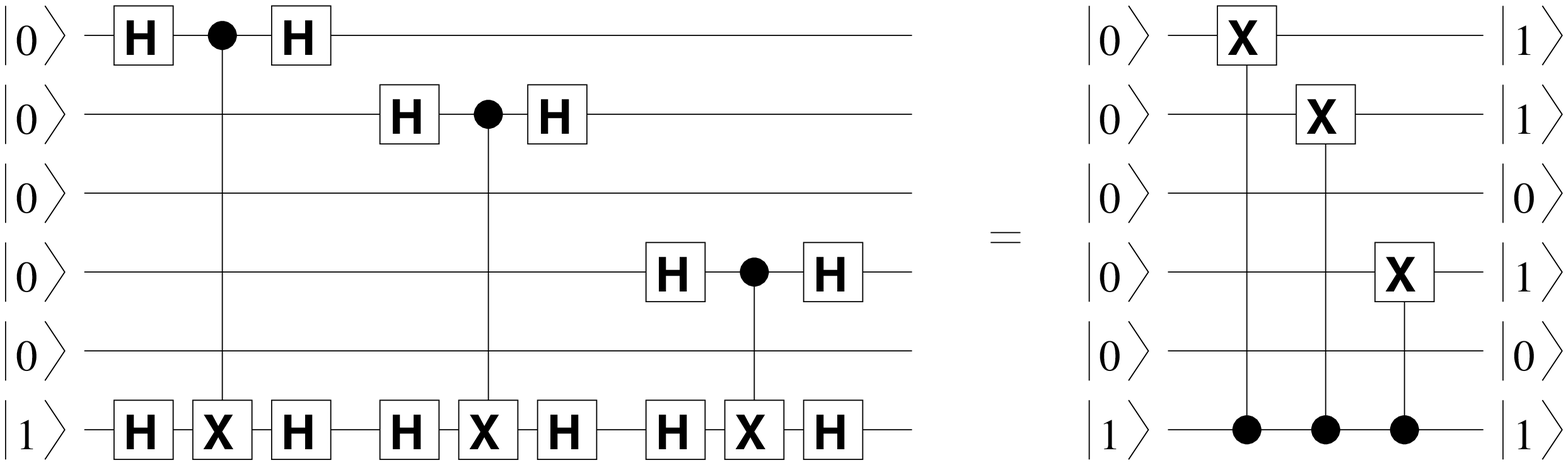}}
\vskip 10pt
{\narrower

\noindent Figure 6.  Because $\Uc_a$ behaves as a collection of cNOT
gates, because Hadamards reverse the action of
cNOT gates, and because the output register has been set to 1, we have
a simple explanation, requiring only one universe, for why $a$ can be
determined with only one invocation of $\Uc_a$.

}
\endinsert 
\vskip 10pt

Notice that the only way quantum mechanics enters is through the
ability of Hada\-mards to reverse the action of cNOT gates, as
illustrated in Figure 4.  Since the same trick can be done classically
with 2-Qbit SWAP gates, as shown in Figure 7, the magic of quantum
mechanics here lies entirely in the possibility it offers for
reversing the roles of target and control Qbits using only 1-Qbit
local operations.  (If the six pairs of vertically separated Hadamards
in the middle circuit of Figure 6 were vertically linked into
irreducibly 2-Qbit gates, then they could no longer be moved to the
extreme right and left of the circuit without leaving any traces in
the central part, as in the upper half of Figure 6.)

\vfil\eject
\hrule
\midinsert 
\epsfxsize=2.4truein
\centerline{\epsfbox{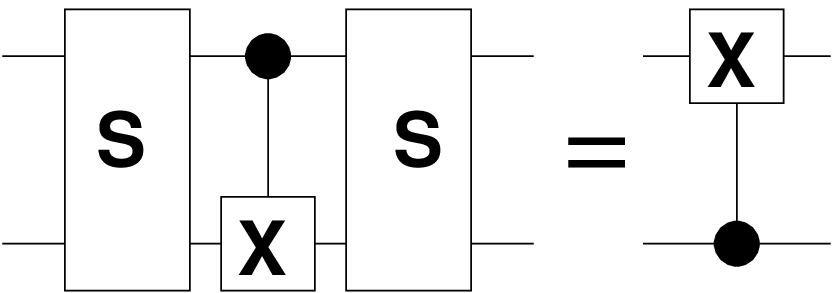}}
\endinsert 

{\narrower

\noindent Figure 7.  Classical SWAP gates also invert the action of a
cNOT gate just as Hadamards do (Figure 4).  The power of Hadamards
over SWAPs is that the Hadamards can do it without requiring any
interaction between the two Qbits.  If swap gates are used to invert
the cNOT gates in the final form of Figure 6, they get all tangled up
with each other when one attempts move them to the edges of the
figure, as one can do with the Hadamards.

}
\vskip 10pt
\hrule

\vfil\eject
\noindent{{\bf 3. How I invented the Copenhagen interpretation while
teaching quantum mechanics
to computer scientists.}}

\vskip 10pt
\hskip 85pt \hbox{\vbox{\baselineskip = 8pt \obeylines 
{\sl [I]n our description of nature the purpose 
is not to disclose the real essence of the phenomena 
but only to track down, so far as it is possible, 
relations between the manifold aspects of our experience.}}

}

\hskip 190pt {\it --- Niels Bohr} [7]
\vskip 15pt

For the past few years I have taught a course in quantum computation
suitable for computer scientists having no background in physics [8].  My
first challenge was to develop a minimalist introduction to quantum
mechanics that straightforwardly conveyed in a few lectures everything
a mathematically sophisticated student needed to know to understand
discussions like, for example, that of the preceding section.

The advantage of teaching an approach to a subject as you develop it
is that you get striking demonstrations of the ways in which it does
and doesn't work.  After several iterations and countless revisions,
reorganizations, and reconstructions, the process seemed to be
converging.  It was only then that I realized that the unproblematic,
no-nonsense, lucid, practical pedagogical approach that had so
painfully evolved out of my clumsy initial attempts, was nothing but
the standard Copenhagen interpretation.  What follows, therefore, is
my vision of why quantum computation, far from demonstrating the
existence of the multiverse, provides the simplest and most compelling
example of a major application of quantum mechanics which the
Copenhagen point of view fits like a glove.

We begin with a silly formulation of ordinary non-quantum {\it
classical\/} computation, based on representing the integers less than
$N$ as orthonormal vectors in $N$ dimensions: $$ 0 \ra
\pmatrix{1\cr0\cr0\cr0\cr0\cr\vdots},\ \ \ 1 \ra
\pmatrix{0\cr1\cr0\cr0\cr0\cr\vdots},\ \ \ 2 \ra
\pmatrix{0\cr0\cr1\cr0\cr0\cr\vdots},\ \ \ 3 \ra
\pmatrix{0\cr0\cr0\cr1\cr0\cr\vdots},\ \ \ 4\ra
\pmatrix{0\cr0\cr0\cr0\cr1\cr\vdots}, \ldots\,. \eq(unary)$$ This
clumsy form takes on a rather simpler structure if $N$ is a power of
2, so we specialize to the case $N = 2^n$.  When $n=1$ we have only
two such vectors, which we denote by a more compact pair of symbols
due to Dirac: $$\pmatrix{1\cr0} = \k0,\ \ \ \pmatrix{0\cr1} =
\k1.\eq(n=1)$$ To manipulate these two numbers in a computer it is
necessary to represent then by a physical system having two
distinguishable configurations.  Continuing to follow Dirac, we call
any such physical system a {\it Cbit\/} [9] (``C'' for ``classical'').
The vectors $\k0$ and $\k1$ associated with these two configurations
are called the {\it states\/} of the Cbit.

If we have two Cbits ($n=2$) their four states conveniently decompose
into the tensor product of two 1-Cbit states: $$ \k0_2 =
\pmatrix{1\cr0\cr0\cr0} = \pmatrix{1\cr0}\x\pmatrix{1\cr0} = \k0\k0 =
\k{00},$$ $$ \k1_2 = \pmatrix{0\cr1\cr0\cr0} =
\pmatrix{1\cr0}\x\pmatrix{0\cr1} = \k0\k1 = \k{01},$$ $$ \k2_2 =
\pmatrix{0\cr0\cr1\cr0} = \pmatrix{0\cr1}\x\pmatrix{1\cr0} = \k1\k0 =
\k{10},$$ $$ \k3_2 = \pmatrix{0\cr0\cr0\cr1} =
\pmatrix{0\cr1}\x\pmatrix{0\cr1} = \k1\k1 = \k{11}.\eq(2cbit)$$ The
last two forms in each line provide some simpler notations for these
2-Cbit states.  Pause to admire how the quantum mechanical practice of
representing the states of composite systems by the tensor product of
the subsystem states emerges automatically from the trivial
representation of integers introduced in \(unary).
  
The tensor product extends straightforwardly to many Cbits: states of
$n$ Cbits can be expressed as tensor products of $n$ 1-Cbit states.
For example, $$\k5_3 = \pmatrix{0\cr 0\cr 0\cr 0\cr 0\cr 1\cr 0\cr
0\cr} = \pmatrix{0\cr 1}\x\pmatrix{1\cr 0}\x\pmatrix{0\cr 1} =
\k1\k0\k1 = \k{101}.\eq(3cbit) $$

While the operation $\Xc$ defined in \(Zc) makes perfect
sense for Cbits (representing the logical NOT) the operation $\Zc$
makes no sense at all, since we have assigned no meaning to the sign
of the state-vector that describes a Cbit.  Nevertheless, combinations of
operators $\Zc$ on pairs of Cbits can be classically meaningful.  For
example 
$$\h\bigl(\oc+\Zc\x\Zc\bigr)\ {\rm projects\ on}\ \k0\k0,\ \ \k1\k1,$$
$$\h\bigl(\oc-\Zc\x\Zc\bigr)\ {\rm projects\ on}\ \k0\k1,\ \ \k1\k0.\eq(proj)$$
This leads directly to a representation of the SWAP operator \Sc\
that takes the 2-Cbit state $\k x\k y$ to $\k y\k x$: 
$$\eqalign{\Sc =&  \h\bigl(\oc+\Zc\x\Zc\bigr)
                        +\bigl(\Xc\x\Xc\bigr)\h\bigl(\oc-\Zc\x\Zc\bigr)\cr
 =&  \h\bigl(\oc+\Zc\x\Zc
           +\Xc\x\Xc - \Yc\x\Yc\bigr),\ \ \ \ \Yc = \Xc\Zc.}\eq(exchg)$$

Pause also to admire the simplicity of this classical derivation of
the form of the quantum-mechanical exchange operator, compared with
the standard derivation based on angular momentum technology.  Note
also the further simplicity introduced into \(exchg) by incorporating
an additional factor of $i$ into the definition of $\Yc$ (which also
makes it hermitian, like $\Xc$ and $\Zc$.)  With
examples like this one can motivate the utility of extending the
notion of states to include multiplication by complex scalars, leading
to the generalization from Cbits to {\it Qbits\/}.

Qbits are physical systems characterized by states which fully exploit the
entire $2^n$ dimensional complex vector space spanned by the $2^n$
orthonormal Cbit states.  Nature has been kind enough to present us
with many examples of them.  The general state $\k\Psi$ of $n$ Qbits is
{\it any\/} unit vector:
$$\k\Psi = \sum_{0\leq x<2^n}a_x\k x_n,\ \ \
\sum_{0\leq x<2^n}|a_x|^2=1.\eq(Psi)$$

With one (extremely important) exception all operations on Qbits are
reversible.  Since the exception (``measurement'') has no nontrivial
analog for Cbits, in comparing Qbits and Cbits it suffices to
consider only reversible operations on Cbits.  The only reversible
operations on the $2^n$ Cbit states are their $(2^n)!$ possible
permutations.  But the general operation nature allows us to perform
on $n$-Qbit states is {\it any\/} linear norm-preserving
transformation,  $$\k\Psi \ra \Uc\k\Psi,\ \ \ \Uc\ {\rm
unitary},\eq(unitary)$$ as shown schematically in Figure 8.

\midinsert 
\hrule
\vskip 10pt
\epsfxsize=2.4truein
\centerline{\epsfbox{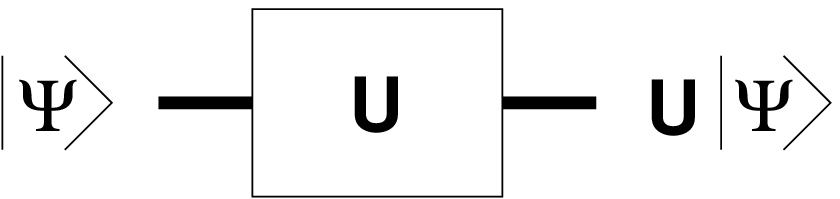}}
{\narrower

\noindent Figure 8. Circuit diagram representing a unitary gate
$\Uc$ acting on $n$ Qbits.

} \vskip 10pt \hrule 
\endinsert 

\vskip 10pt  

While Qbits are far more versatile than Cbits in their range of states
and the operations one can perform on them, the usefulness of their
versatility is highly constrained by one important difference between
Qbits and Cbits.  Learning the state $\k x_n$ of $n$ Cbits is
unproblematic: you just look to see which of the $2^n$ possible states
$\k x_n$ it is.  In contrast, learning the state $\k\Psi_n =
\sum_x a_x\k x_n$ associated with $n$ Qbits is impossible.  Given the
Qbits there is nothing you can do to them to reveal their state.

To extract any information from Qbits one must ``make a
measurement''.  This consists of sending the Qbits through a ``measurement
gate''.  If the state of the Qbits is \(Psi) then the measurement gate
signals $x$ with probability $p=|a_x|^2$. After $x$ is signaled, the
state associated with the Qbits must be taken to be $\k x_n$.  These
features of an $n$-Qbit measurement gate and the fact that it
represents the only way to extract information from the $n$-Qbits is
known as the {\it Born rule\/}.  The Born rule is illustrated in
Figure 9.  \vskip 10pt \hrule

\midinsert 
\epsfxsize=2truein
\centerline{\epsfbox{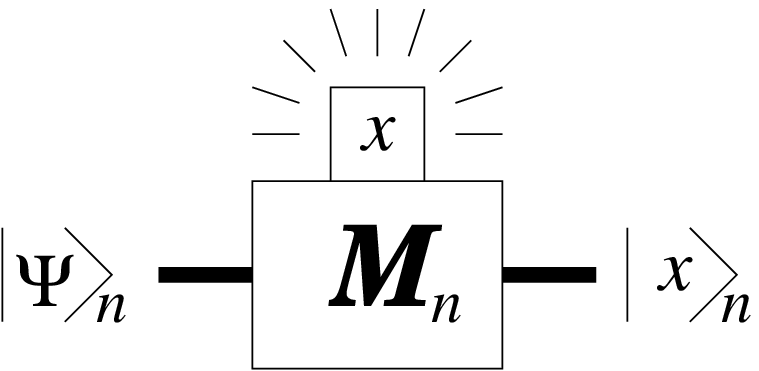}}
\endinsert 
{\narrower

\noindent Figure 9. Circuit diagram representing an $n$-Qbit
measurement gate.

} \vskip 10pt \hrule \vskip 10pt

As defined above ``measurement" is always in the
computational basis.  One loses nothing by this simplification, since
measurement in any other basis can be described as measurement in the
computational basis, preceded by an appropriate unitary transformation.
What one gains, pedagogically, is the need to invoke only a single
variety of measurement gate and, as noted below, only 1-Qbit
measurement gates.

A somewhat stronger version of the Born rule plays a crucial role in
quantum computation, though it is rarely explicitly mentioned in most
standard quantum mechanics texts.  The stronger form applies when one
measures only a single one of $n$ Qbits.  The state $\k\Psi$ of all
$n$ Qbits can always be represented in the form $$\k\Psi =
a_0\k0\k{\Phi_0} + a_1\k1\k{\Phi_1},\ \ \ \ |a_0|^2 + |a_1|^2 =
1,\eq(genborn)$$ where the Qbit to be measured appears on the left and
where $\k{\Phi_0}$ and $\k{\Phi_1}$ are normalized but not necessarily
orthogonal states of the $n-1$ unmeasured Qbits.  The generalized Born
rule asserts that if only the single Qbit is measured then the 1-Qbit
measurement gate will indicate $x$ (0 or 1) with probability
$|a_x|^2$, after which the $n$-Qbit state will be the product state
$\k x\k{\Phi_x}$, as illustrated in Figure 10.
\midinsert 
\hrule
\vskip 10pt
\epsfxsize=5truein
\centerline{\epsfbox{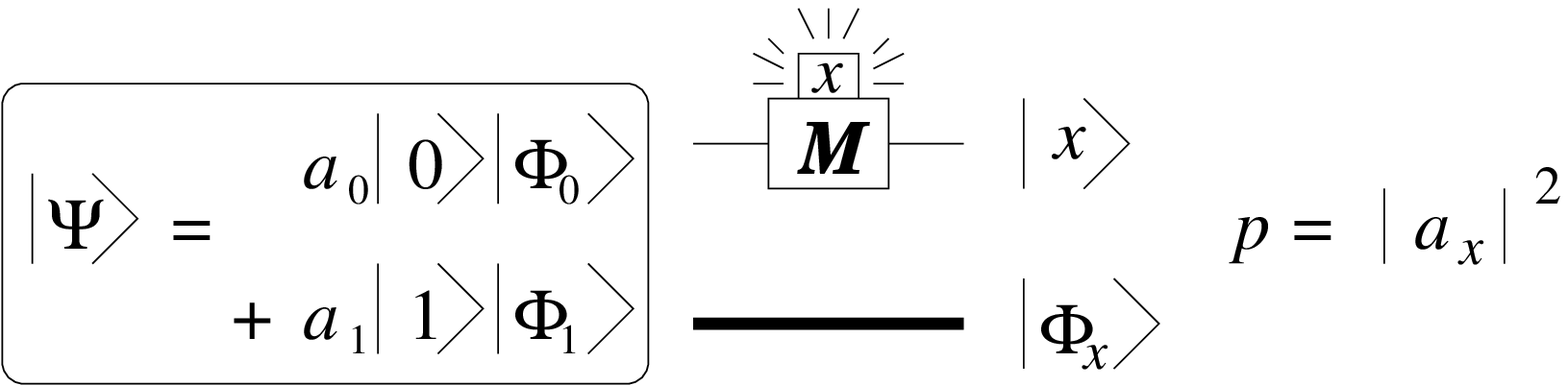}}
\vskip 8pt
{\narrower

\noindent Figure 10.  Action of a 1-Qbit measurement gate on a single
one of $n$ Qbits, according to the generalized Born rule.

}
\vskip 10pt
\hrule
\endinsert

To see that the gate acting on the measured Qbit in Figure 10 is
indeed the $n=1$ version of the $n$-Qbit measurement gate of Figure 9,
note that Figure 10 simplifies to Figure 11 in the special case
$\k{\Phi_0} = \k{\Phi_1} = \k{\Phi}$.  Now the entangled input state
in Figure 10 becomes an uncorrelated product in which both the
measured Qbit and the remaining $n-1$ Qbits have states of their own.
The $(n-1)$ unmeasured Qbits now take no part whatever in the process.
Nothing acts on them and they do nothing but maintain their original
state $\k\Phi$.  Their presence is irrelevant to the upper part of the
figure, which is nothing more than the $n=1$ version of Figure 10.

\vskip 10pt \hrule \midinsert \epsfxsize=4.5truein
\centerline{\epsfbox{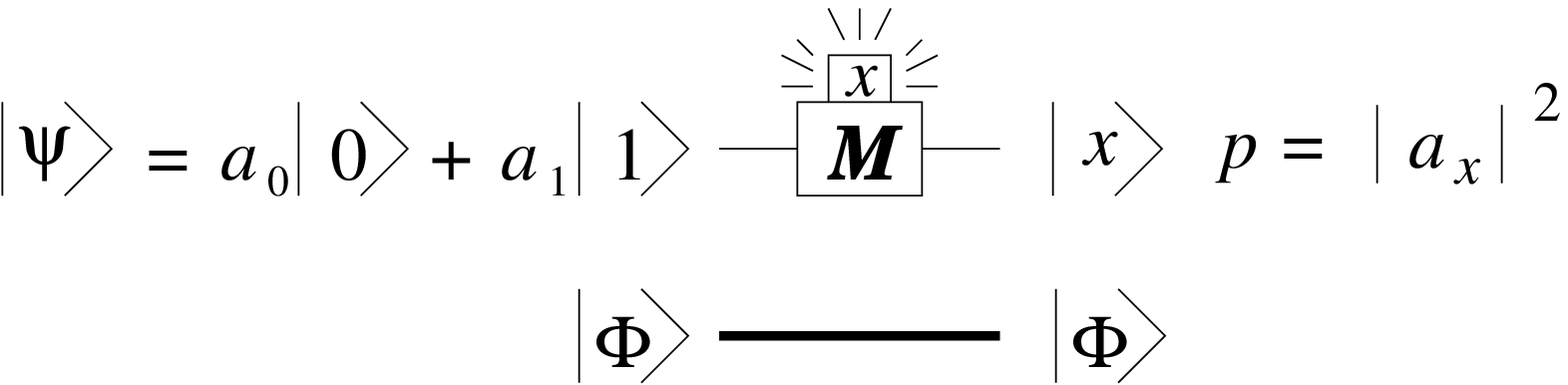}} \endinsert {\narrower \noindent
Figure 11.  Simplification of Figure 10 when $\k{\Phi_0} = \k{\Phi_1}
= \k{\Phi}$

 } 

\vskip 10pt

\hrule

\vskip 10pt

It is an elementary consequence of the generalized Born rule that the
$n$-Qbit measurement gate of the ordinary Born rule can be constructed
out of $n$ 1-Qbit measurement gates, as illustrated in Figure 12.
\midinsert

\hrule
\vskip 10pt
\epsfxsize=2.5truein
\centerline{\epsfbox{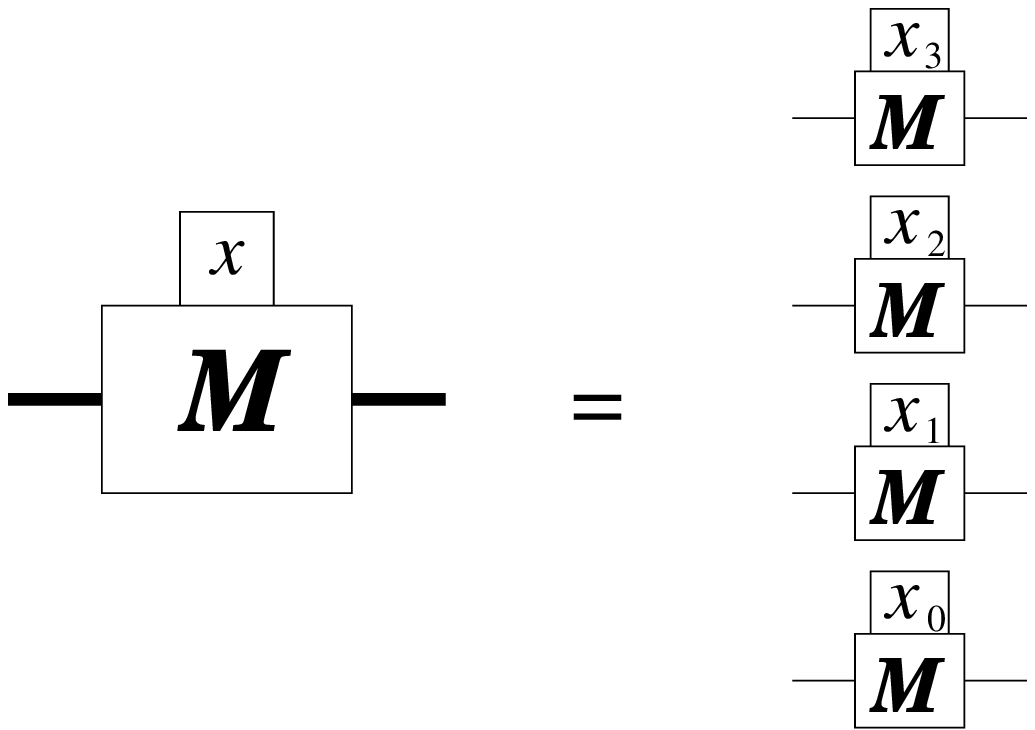}}
{\narrower

\noindent Figure 12.  Constructing a 4-Qbit measurement gate out of
four 1-Qbit measurement gates.  The integer $x$ has the binary
expansion $x_3x_2x_1x_0$.

}

\vskip 10pt
\hrule
\endinsert 

Although the generalized Born rule is stronger, it follows from the
Born rule under two plausible assumptions.  (a) Once a Qbit ceases to
interact with others and ceases to be acted on by unitary gates it
doesn't matter when it is measured. (b) To assign a state to Qbits is
to do nothing more than to specify the probabilities of subsequent
measurement outcomes, possibly preceded by unitary gates.  Since the
generalized Born rule reduces the notion of measurement to a single
black-boxed 1-Qbit measurement gate (and indeed, since Qbits can be
measured one by one, one needs only a single specimen of such a
measurement gate) by far the most economical introduction to quantum
computation is to base it on a primitive concept of the 1-Qbit
measurement gate and make explicit assumptions (a) and (b) above.

Since it is impossible to determine the state of a collection of Qbits
from the Qbits themselves, how is one to associate with the registers
of a quantum computer the initial states on which the unitary
transformations subsequently act?  The obvious, simplest, and
conceptually most economical answer is to exploit the measurement
gates themselves.  One can initialize $n$ Qbits to the state $\k0_n$
by measuring each Qbit and applying $\Xc$ if and only if the
measurement indicates 1, as illustrated in Figure 13.  \vskip 8pt
\midinsert \hrule \vskip 8pt \epsfxsize=2.0truein
\centerline{\epsfbox{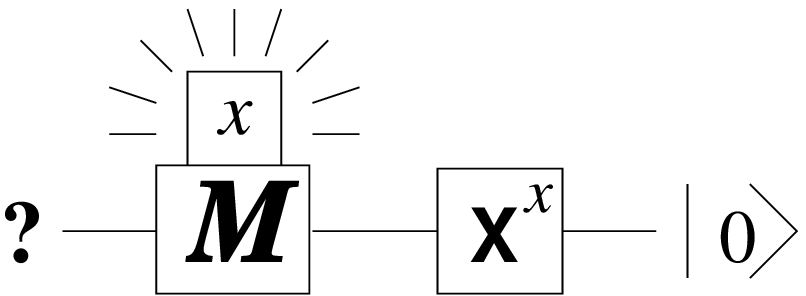}} {\narrower

\noindent Figure 13.  How to use a 1-Qbit measurement gate to prepare
an off-the-shelf Qbit so its associated state is $\k0$.

}

\vskip 10pt
\hrule
\endinsert 

Note the following features of this pedagogically motivated approach
to quantum mechanics:

(1) It relies on an irreducible primitive notion of a unique
black-boxed 1-Qbit measurement gate.  The measurement gate is the only
irreversible circuit element.  It is defined by what it does, and what
it does is to extract information from the Qbits in a form that is
immediately accessible to {\it us}.  There is no other way for us to
get such information from the Qbits [10].

(2) Measurement plays a dual role.  Output on a readable (classical)
display not only ends the computation.  It also provides, without any
further complication, a straightforward way to {\it begin\/} the
computation. Initialize every Qbit to the 1-Qbit state $\k0$ by
sending each through a measurement gate, and then do nothing or
apply $\Xc$, depending on whether the display shows 0 or 1.

(3) Unlike the state of $n$ Cbits, the state of $n$ Qbits does not
reside in the Qbits themselves: presented with a bunch of Qbits there
is nothing one can do to them to reveal their state.  Indeed, in
general --- for example if they share with other Qbits an entangled
state --- Qbits will not have a state of their own at all.  To
determine the state of Qbits (or whether they have one) one must go
ask Alice, who knows their history: what initial measurements were
performed on them, what the outcomes of the initial measurements were,
and what subsequent unitary gates were applied.

(4) While the {\it purpose\/} of the state of $n$ Cbits is an
anthropocentric add-on to its intrinsic character (what, for example,
is the purpose of the velocity of a classical particle?) one would not
bother to keep track of the state of $n$ Qbits were it not that this
information about their past history has a specific purpose: it
enables us to determine the correlations between the initial and final
measurement outcomes after any intermediate sequence of applications
of unitary gates.

All of these features resonate strongly with the constellation of
ideas known as the Copenhagen interpretation.  The quantum state of a
system is not an objective property of that system.  It merely provides
an algorithm enabling one to infer from the initial set of measurements
and outcomes (``state preparation'') the probabilities of the results
of a final set of measurements after a specified intermediate time
evolution.  We ourselves have direct access to nothing beyond the
outcomes of such measurements.

Why did my bare-bones, no-nonsense, pedagogically motivated,
minimalist introduction to quantum mechanics come out sounding so
Copenhagen?  I think there are several reasons:

(a) Proponents of the Copenhagen interpretation (notably Heisenberg
and Peierls) have always maintained that the quantum mechanical
formalism does not describe ``the system'' but ``our knowledge of the
system''.  Quantum computation is the first application of quantum
mechanics that does not use it to further our understanding of the
physical world.  On the contrary, quantum computation exploits the
known quantum mechanical character of the physical world to expedite
the processing of knowledge, as represented symbolically by
constituents (Qbits) of that world.  It is therefore not surprising
that the Copenhagen interpretation should provide a congenial setting
for the exposition of quantum computation.

(b) A computation uses a finite set of Qbits.  It has an unambiguous
beginning and end.  There is always a world external to the
computation.  If there were not an outside world there would be no
point in doing the computation because there would be nobody or
nothing to take advantage of the output.  Nobody (well, at this stage
practically nobody) wants to view the entire universe (single or
multiple) as one colossal quantum computer, sufficient unto itself.
The Copenhagen interpretation is characterized by a similar modesty of
scope.  Physics is a tool for relating some aspects of our experience to
other aspects.  Every application of physics begins and ends with an
appeal to experience.

(c) The pedagogical device of restricting ``measurement'' to
measurement in the computational basis, treating measurement in other
bases as computational-basis measurement preceded by an appropriate
unitary transformation, resonates with the Copenhagen notion of the
primacy of the classical world.  The computational basis states are
(actually one should replace ``are'' by ``are isomorphic to'') the
states that describe ordinary classical Cbits.  By restricting
``measurement'' to the computational basis I have automatically
arranged for the input and output of every quantum computation to be
describable in the ordinary old-fashioned language of classical
computation --- numbers flashed on a classical display.  Bohr always
insisted that our knowledge of the world had to be formulated in
ordinary language, or we could not communicate it to anybody else.

(d) My computer science students know very little physics.  They are
therefore immune to any temptation to reify the states of Qbits into
properties of the associated physical systems.  If you think you too
are immune from such temptation, ask yourself whether you do or do not
belief that a horizontally polarized photon is {\it intrinsically\/}
different from a vertically polarized photon.  If you do, you are a
victim of that very temptation.

I conclude by translating the possibly obscure quotation at
the head of this section into the quite straightforward form it assumes
in the context of quantum computation:

{\sl In our description of a quantum computation the purpose is not to
disclose the real essence of the Qbits but only to track down
statistical relations between initial and final measurement outcomes.}

\bigskip

\vfil\eject

\noindent {\bf Acknowledgment}

Supported by the National Science Foundation under grant No.~0098429.  

\bigskip

\noindent {\bf References}

\noindent 1.  D.~DiVincenzo, P.~Hayden, and B.~Terhal,  
``Hiding quantum data'', {\tt www.arxiv.org/
abs/quant-ph/0207147}.
\vskip 4pt

\noindent 2. C.~H.~Bennett et al., ``Teleporting an unknown quantum
state via dual classical and Einstein-Podolsky-Rosen channels'',
Phys.~Rev.~Lett.~{\bf 70}, 1895-99 (1993).
\vskip 4pt

\noindent 3. C.~A.~Fuchs, ``Notes on a Paulian idea: foundational, historical,
   anecdotal and forward-looking thoughts on the quantum'', 
 {\tt www.arxiv.org/abs/quant-ph/0105039}.
\vskip 4pt

\noindent 4. C.~H.~Bennet and S.~J.~Wiesner, ``Communication via one-
and two-particle operators on Einstein-Podolsky-Rosen states'',
Phys. Rev. Lett. {\bf 69}, 2881-84 (1992).
\vskip 4pt

\noindent 5. N.~D.~Mermin, ``Deconstructing dense coding'', Phys.~Rev.~A {\bf
66}, 032308 (2002).
\vskip 4pt

\noindent 6. I use here the unorthodox spelling {\it Qbit\/} because
   it will be
   constantly juxtaposed to {\it Cbit\/}, a role the currently favored
   {\it qubit\/} cannot gracefully play.

\vskip 4pt
\noindent 7. N.~Bohr, {\it Collected Works} 
vol.~6, p.~296, North Holland (1985). 
\vskip 4pt

\vskip 4pt
\noindent 8. The pedagogical approach to quantum mechanics summarized
here is described in more detail in N.~D.~Mermin, ``From Cbits to
Qbits: teaching computer scientists quantum mechanics'', American Journal
of Physics {\bf 71}, 23-30 (2003).  Lecture notes for the course
itself are regularly updated, revised, and reposted at 
{\tt http://www.ccmr.cornell
.edu/\~{}mermin/qcomp/CS483.html}.

\vskip 4pt \noindent 9. The term ``c-bit'' doesn't work because one
often needs to talk, for example, about 2-Cbit states.

\vskip 4pt \noindent 10.  The art of quantum computation, of course,
is to construct unitary transformations leading to final states in
which only informative values of $x$ are associated with appreciable
probabilities $|a_x|^2$.

\bye